\documentstyle[12pt,epsf,captions]{article}
\def\singlespace 
{\smallskipamount=3.75pt plus1pt minus1pt
\medskipamount=7.5pt plus2pt minus2pt
\bigskipamount=15pa plus4pa minus4pt \normalbaselineskip=12pt plus0pt
minus0pt \normallineskip=1pt \normallineskiplimit=0pt \jot=3.75pt
{\def\smallskip {\vskip\smallskipamount}} {\def\medskip
{\vskip\medskipamount}} {\def\bigskip {\vskip\bigskipamount}}
{\setbox\strutbox=\hbox{\vrule height10.5pt depth4.5pt width 0pt}}
\parskip 7.5pt \normalbaselines} 

\def\middlespace
{\smallskipamount=5.625pt plus1.5pt minus1.5pt \medskipamount=11.25pt
plus3pt minus3pt \bigskipamount=22.5pt plus6pt minus6pt
\normalbaselineskip=22.5pt plus0pt minus0pt \normallineskip=1pt
\normallineskiplimit=0pt \jot=5.625pt {\def\smallskip
{\vskip\smallskipamount}} {\def\medskip {\vskip\medskipamount}}
{\def\bigskip {\vskip\bigskipamount}} {\setbox\strutbox=\hbox{\vrule
height15.75pt depth6.75pt width 0pt}} \parskip 11.25pt
\normalbaselines} 

\def\doublespace 
{\smallskipamount=7.5pt plus2pt minus2pt \medskipamount=15pt plus4pt
minus4pt \bigskipamount=30pt plus8pt minus8pt \normalbaselineskip=30pt
plus0pt minus0pt
\normallineskip=2pt \normallineskiplimit=0pt \jot=7.5pt
{\def\smallskip {\vskip\smallskipamount}} {\def\medskip
{\vskip\medskipamount}} {\def\bigskip {\vskip\bigskipamount}}
{\setbox\strutbox=\hbox{\vrule height21.0pt depth9.0pt width 0pt}}
\parskip 15.0pt \normalbaselines}

\def\lpp{\lambda^{\prime \prime}}
\def\lbf{\lambda^{\prime \prime}_{133}}
\def\lbs{\lambda^{\prime \prime}_{233}}
\def\lbv{\lambda^{\prime \prime}_{B\!\!\!/}}

\def\abv{A^0_{B\!\!\!/}}
\newcommand{\be}{\begin{equation}}
\newcommand{\ee}{\end{equation}}
\newcommand{\bea}{\begin{eqnarray}}
\newcommand{\eea}{\end{eqnarray}}
\begin{document}
\middlespace
\vskip 2cm
\begin{center}
\large {\bf Signatures of Baryon non-conserving Yukawa couplings in a 
supersymmetric theory} \\ 
\vskip 1in Mar Bastero-Gil$^{a}$ and Biswajoy Brahmachari$^{b}$ \\
\end{center}
\begin{center}

(a) Scuola Internazionale Superiore di Studi Avanzati \\ 34013
Trieste, ITALY. \\ (b) International Centre For Theoretical Physics,\\
34100 Trieste, ITALY.\\
\end{center}
\vskip 1in
{
\begin{center}
\underbar{Abstract} \\
\end{center}

Renormalization effects of large baryon-nonconserving Yukawa
couplings $\lambda^{\prime\prime}_{ijk} \overline{U_i} \overline{D_j}
\overline{D_k}$ lower the right handed squark masses keeping the left-handed
squark masses virtually untouched at the lowest order. At low energy
they enhance the mass-splitting between left and right handed squarks of 
the same generation as well as intergenerational mass splitting among 
squarks, potentially detectable in future  colliders or in rare 
decays. The predicted mass of the lightest stop squark becomes 
lower than the experimental bound for larger ranges of parameter space 
than that of the Baryon-conserving case, hence, further constraining the 
parameter space of a supersymmetric theory when baryon violation is included.}

\newpage
\doublespace 
We know that the radiative corrections due to a large top quark Yukawa
coupling drives a squared Higgs scalar mass to negative values
triggering the radiative electroweak symmetry breaking \cite{ewbr} in
the Minimal Supersymmetric Standard Model (MSSM). Can similar effects
be caused by other large Yukawa couplings? In particular, MSSM allows
the presence of lepton and baryon number violating couplings
\cite{rpty} unless they are forbidden by fiat invoking R-parity. In
the models with explicit R-parity violation some of these couplings,
especially the ones involving the third generation fields, are almost
unconstrained by experiments. They could as well be as large as the
top quark Yukawa coupling. Such Yukawa couplings may involve (for
example) colored fields. Hence, intuitively we may expect that the
radiative corrections due to such potentially large couplings can have
analogous effects on squark masses, thereby serving as an indirect
signature of R-parity violation. Here we explore this possibility.

Apart from the standard Yukawa couplings related to the fermion
masses, Gauge invariance and Lorentz invariance allows the R-parity
violating Yukawa couplings in the low energy effective supersymmetric
theory, given by,
\begin{eqnarray}
W_{R\!\!\!/} = {\lambda_{ijk} \over 2 } L_i L_j E^c_k+
\lambda^{\prime}_{ijk} L_i Q_j D^c_k + {\lambda^{\prime \prime}_{ijk}
\over 2} D^c_i D^c_j U^c_k.
\end{eqnarray}
The simultaneous presence of the lepton and baryon number violating
couplings give rise to fast proton decay unobserved in nature
\cite{vissm}. This leaves us the choice of allowing either the baryon
number violating couplings ($\lambda^{\prime \prime}$) or the lepton
number violating couplings ($\lambda$, $\lambda^{\prime}$)
\cite{comment}. There are existing 
experimental bounds on most of the lepton number violating couplings
\cite{lv}, and so, we will consider the first scenario allowing only
the presence of $\lpp$ couplings which stand relatively unconstrained
from experimental results as well as cosmological bounds.

The color antisymmetry in the first and second indices of $\lpp_{ijk}$
leaves us nine independent couplings. All of them were believed to
have strong cosmological constraints \cite{rpconst} due to the
conjecture that they would erase the primordial baryon asymmetry
\cite{primordial}.  However, it has been shown \cite{ross} that such
constraints are model dependent and $\lpp$ can be completely free of
cosmological bounds in scenarios with GUT-era leptogenesis. The
laboratory bounds on $\lpp_{1j1}$ are derived from the non-observation
of $N$-$\overline{N}$ oscillation and double nucleon decay processes
\cite{zwi,gs}. The $\lpp$ couplings involving the third 
generation twice, i.e., $\lpp_{133}$ and $\lpp_{233}$ are bounded from
above by perturbative unitarity arguments 
\cite{pertur} and they can very well be of order O(1) at the 
unification scale. Moreover, we know that the third generation Yukawa
couplings in the R-conserving sector are the largest ones; keeping
this in mind we will have a similar hypothesis in the R-violating
sector. We will consider the effects of $\lpp_{133}$ and $\lpp_{233}$,
where the third generation occurs twice, discarding the rest, on the
renormalization of soft mass terms here.

The theoretical upper bounds \cite{pertur,gs,pertur1} on $\lbf$ and
$\lbs$ can be found demanding the perturbative unitarity of {\it all}
the Yukawa couplings of the theory up to the unification scale $M_X$,
taken to be $2 \times 10^{16}$ GeV here. The essential point to get
these bounds is that when $\lbf$ or $\lbs$ are large enough, they
drive the top quark Yukawa coupling $h_t$ into the non-perturbative
domain before $M_X$, and they may also blow up along with $h_t$ for
low top quark masses, e.g., $m^{pole}_t=165$ GeV in the low $\tan
\beta$ region.  We note that for a fixed $m_t$, the value of $h_t$
depends on the ratio of the vevs of the two Higgs scalars $\tan
\beta=\langle H_2 \rangle / \langle H_1 \rangle$. Hence, the largest
allowed value for $\lbf$ and $\lbs$ depends on $\tan \beta$. We have
taken\footnote{The value of $\alpha_s(m_Z)$ is chosen to have
unification in the gauge sector at the one-loop level.}
$\alpha_s(m_Z)=0.114$, $m_b(m_b)=4.4$ GeV, $m_\tau(m_\tau)=1.777$ GeV
and several values $m^{pole}_t $. Using these values and the
perturbative unitarity argument for {\it all} Yukawa couplings, we get
the upper bounds on,
\be
\lbv= \sqrt{ {\lbf}^2 + {\lbs}^2}.
\ee
These bounds are plotted in Figure (1) as a function of $\tan \beta$.
We have not assumed $b-\tau$ unification to be as model independent as
possible. Figure (1.a) shows the bounds at $M_X$, and Figure (1.b)
those extrapolated to the scale $m_t$.  To see the maximal
renormalization effect on the squark masses for a given $m^{pole}_t$
we have fixed $\tan \beta$ where $\lbv$ is at its maximum.  For
example, in our case\footnote{This perturbative
bound is on $\lbv$. Individual experimental bounds coming from LEP are 
($\lpp_{ij3} ({m_Z}) \le 0.97$) \cite{gautam}. For the bounds on the 
product  $\lbf \lbs$ see Reference \cite{carlson}.}: 
$m^{pole}_t=175$ GeV, $\tan \beta=14.99$ and
$\lbv(M_X)= 2.62$ ($\lbv(m_t)=1.03$).  { \oddsidemargin -.5cm
\begin{figure}[t]
\begin{tabular}{cc}
\epsfysize=7cm \epsfxsize=7cm \hfil \epsfbox{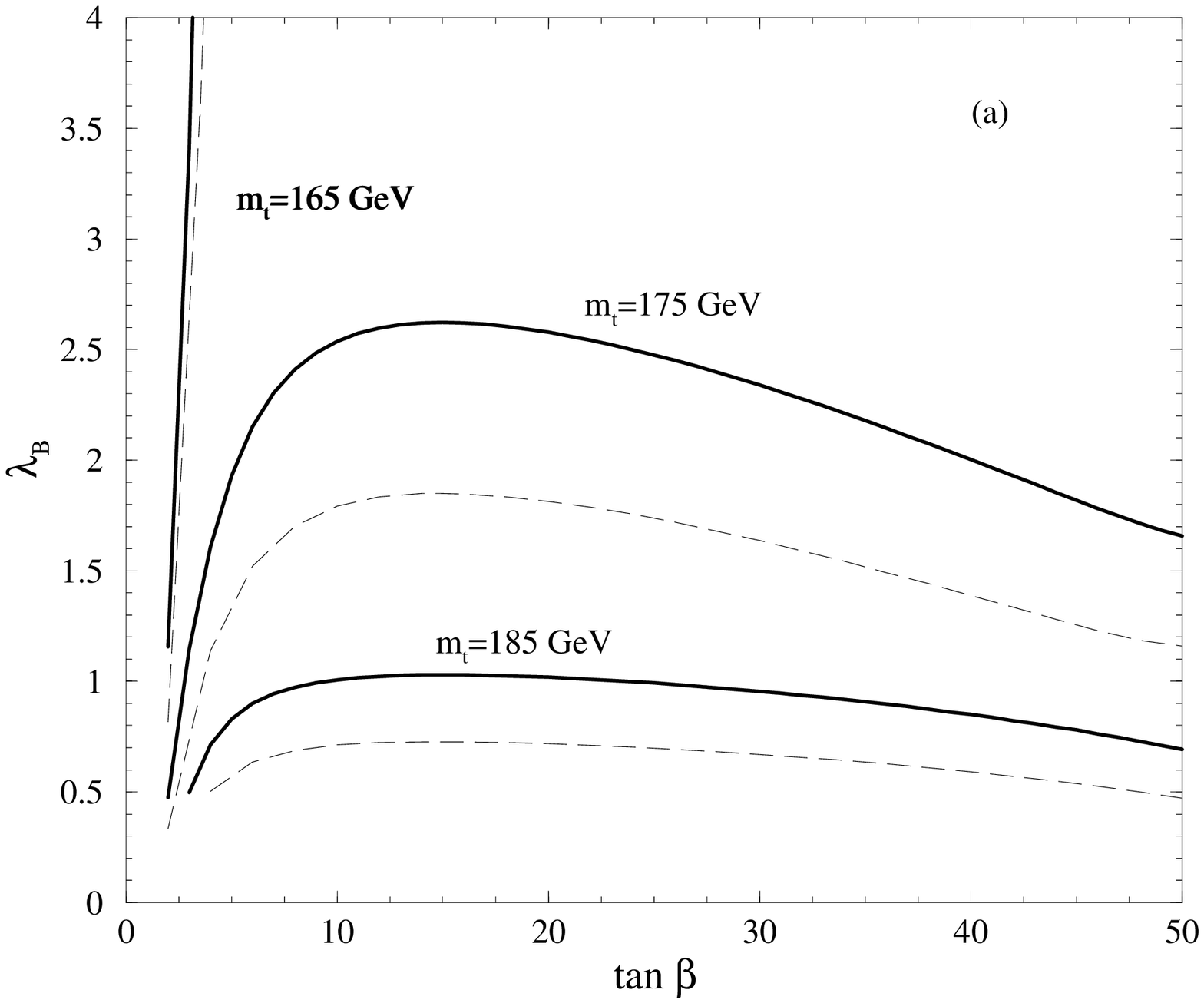} \hfil
& \epsfysize=7cm \epsfxsize=7cm \hfil \epsfbox{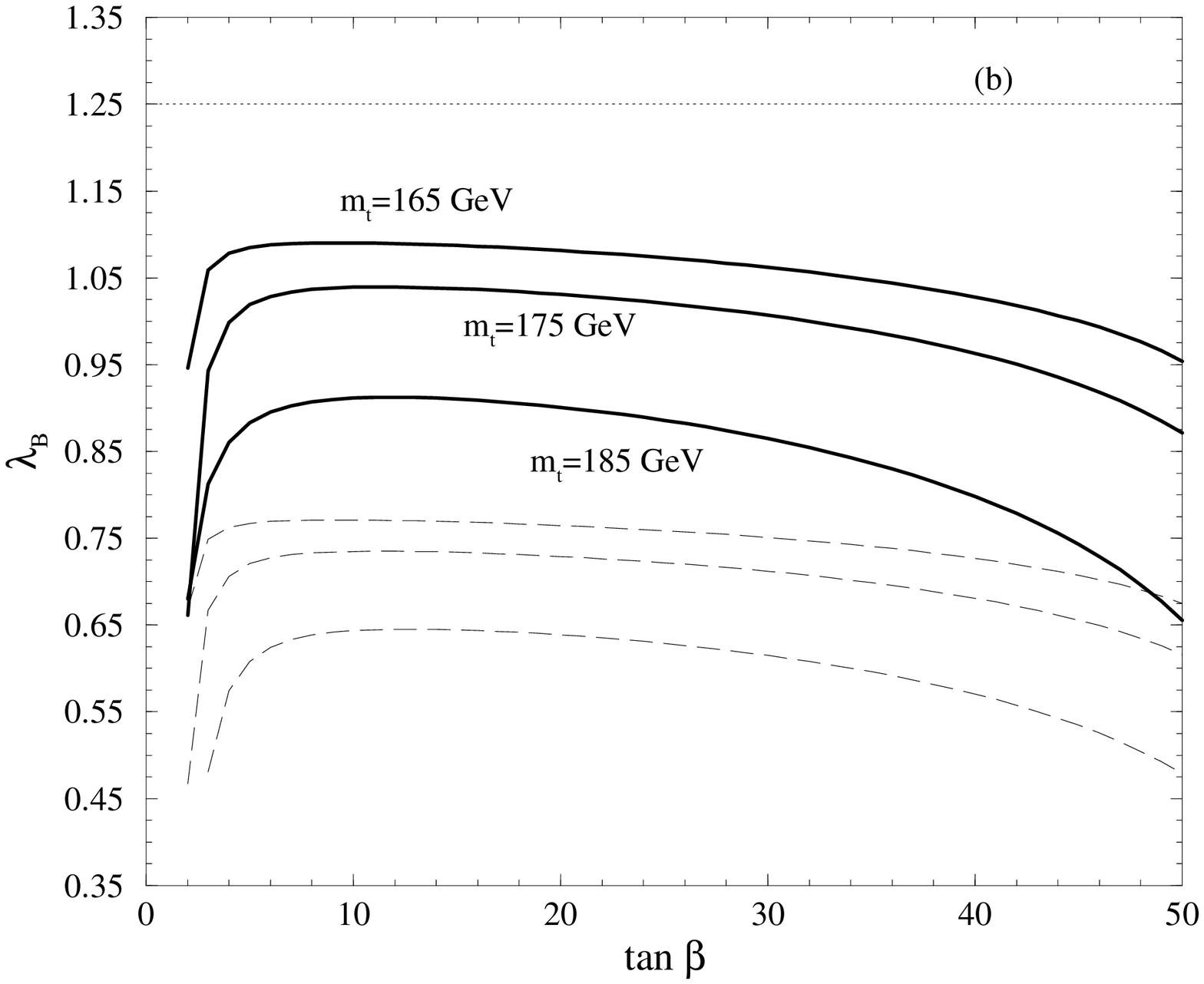} \hfil
\end{tabular}
\captions{Largest allowed values of $\lbv$ (a) at the scale $M_X$ 
and (b) at the scale $m_t$, for $m_b=4.4$ GeV, $m_\tau=1.777$ GeV,
$\alpha_s(m_Z)=0.114$. The dashed lines represent the case where
$\lbf=\lbs$. The dotted horizontal line is the case where $h_t$, $h_b$
and $h_\tau$ are neglected as in Reference [10].
\label{fig1}}
\end{figure}
} We have derived the evolution equations of the soft masses,
including $\lbf$ and $\lbs$ \cite{mar5}. In the case $\lbv=0$ the
equations can be found in Ref. \cite{ewbr}. However, the essential
differences in the evolution equations can be traced from Eqns.
(\ref{splitt}), (\ref{splitb}), (\ref{splits}) given below.  We have
run the coupled set of renormalization group equations (RGE) for the
soft masses and dimensionless Yukawa couplings numerically, taking into 
account the constraints coming from the radiative electroweak symmetry 
breaking. The unknown universal 
soft supersymmetry breaking parameters at the unification scale are,
\[
m_0,~m_{1/2},~A_0,~\abv,~\mu_0,~B_0,
\]
where we have taken $A^0_{133}=A^0_{233}=\abv=A_0$ for simplicity, and
we have chosen the sign of $\mu_0$ to be positive.  We have varied
$m_{1/2}$ in the range 50 to 500 GeV, $m_0$ in the range 100 to 300
GeV, for the cases $A_0=0$ GeV and $A_0=100$ GeV.  The other two
parameters, $\mu_0$ and $B_0$, will be fixed by the choice of $\tan
\beta= 14.99$ and the condition of electroweak symmetry breaking,
given by,
\bea
B(\overline{Q})&=&{\left[{(m^2_{H_1}+m^2_{H_2} + 2 \mu^2) \sin 2 \beta
\over 2 \mu }\right]}_{\overline{Q}},\label{b0} \\
m^2_Z(\overline{Q})&=&2 {\left[{m^2_{H_1}-m^2_{H_2}\tan^2\beta\over
\tan^2 \beta-1} - \mu^2\right]}_{\overline{Q}},
\label{mzq}
\eea 
where $\overline{Q}$ is the scale \cite{qbar} in which the minimum of
the one-loop effective Higgs potential \cite{effective} becomes
identical to the tree level one, given by,
\be
\overline{Q}={1\over\sqrt{e}}\prod_{i}({m_i})^{ d_i m^2_i / 
\sum_{p} d_p m^2_p},
\ee
and,
\be
d_p=(-)^{2s_p}(2s_p + 1){ \partial m^2_p \over \partial v_2}.
\ee
where $m_p$ are $all$ the mass eigenvalues at the scale
$\overline{Q}$. When $\lbv$ is turned on, $h_t(m_t)$ tends to decrease
so $h_t(M_X)$ has to increase; however, the renormalization effects of
the B-violating couplings reduce the right handed squark masses (which
is multiplied by $h_t$ in the $\beta$-function), therefore $m_{H_2}^2$ 
increases
(i.e., it becomes less negative), while $m_{H_1}^2$ remains almost the
same. From Eqn.(\ref{mzq}) we see that for a fixed value of $\tan
\beta$, the value of $\mu(\overline{Q})$ has to decrease. This is
achieved due to the increased top quark Yukawa coupling involved in
the RGE for $\mu$ which gives the correct $h_t(m_t)$ to fit
$m^{pole}_t$ too.  However, in order to avoid too much reduction in
$\mu(\overline{Q})$, the initial value $\mu_0$ has to increase.

Absence of the renormalization effect due to the R-parity violating
couplings in the left handed sector creates an additional splitting
between the squark masses.  In order to get the physical masses of the
squarks after electroweak symmetry breaking, one has to diagonalize
the corresponding mass matrices. For example, in the case of the third
generation up-type squarks the left-right mass matrix is
\cite{effective},
\be
{\cal M}_u=\pmatrix{{M}^2_L & m_u~[A_u + \mu \cot \beta] \cr m_u~[A_u
+ \mu \cot \beta] & M^2_R},
\ee
where,
\be
M^2_{L/R} = {\tilde{m}}^2_{L/R} + m^2_u + ({\rm D-terms})_{L/R}.
\ee
For the down-type squarks $\cot \beta$ is replaced by $\tan \beta$.
For the first and second generation down type squarks, off-diagonal
elements are generated by renormalization effects of $\lbf$ and
$\lbs$, giving the $2 \times 2$ right-right mass matrix,
\be
{{\cal M}^{RR}_{12}}=\pmatrix{{M}^2_{dd} & {\tilde{m}}^2_{ds} \cr
{\tilde{m}}^2_{ds}& M^2_{ss}} \label{dsmat},
\ee
where,
\be
{{\cal M}_{ii}}={\tilde{m}}^2_{iR} + ({\rm D-terms}).
\ee
All the entries in the mass matrices are running parameters.  We have
shown the difference in the eigenvalues of the stop and sbottom squarks
evaluated at their own scale in Figure (2.a). The dashed lines are
for the case $\lbv=0$. The splitting between the squark eigenvalues
increases with the increasing $m_{1/2}$ as well as with increasing
$m_0$. The mass of the lighter stop is plotted in Figure (2.b). We
notice that for large values of $m_0$ and low values of $m_{1/2}$ it
can be below the experimental bound. In Figure (3) we have plotted the
difference in mass eigenvalues of the first and the second generation
right handed squarks.  Changing the value of $A_0$ from 0 to 100 GeV
we have not found any substantial difference in these results with
B-violation shown in the solid lines. However, such a change in $A_0$
has observable effects in the difference in the stop eigenvalues when
$\lbv=0$.  

{ \oddsidemargin -1cm
\begin{figure}[t]
\begin{tabular}{cc}
\epsfysize=7cm \epsfxsize=7cm \hfil \epsfbox{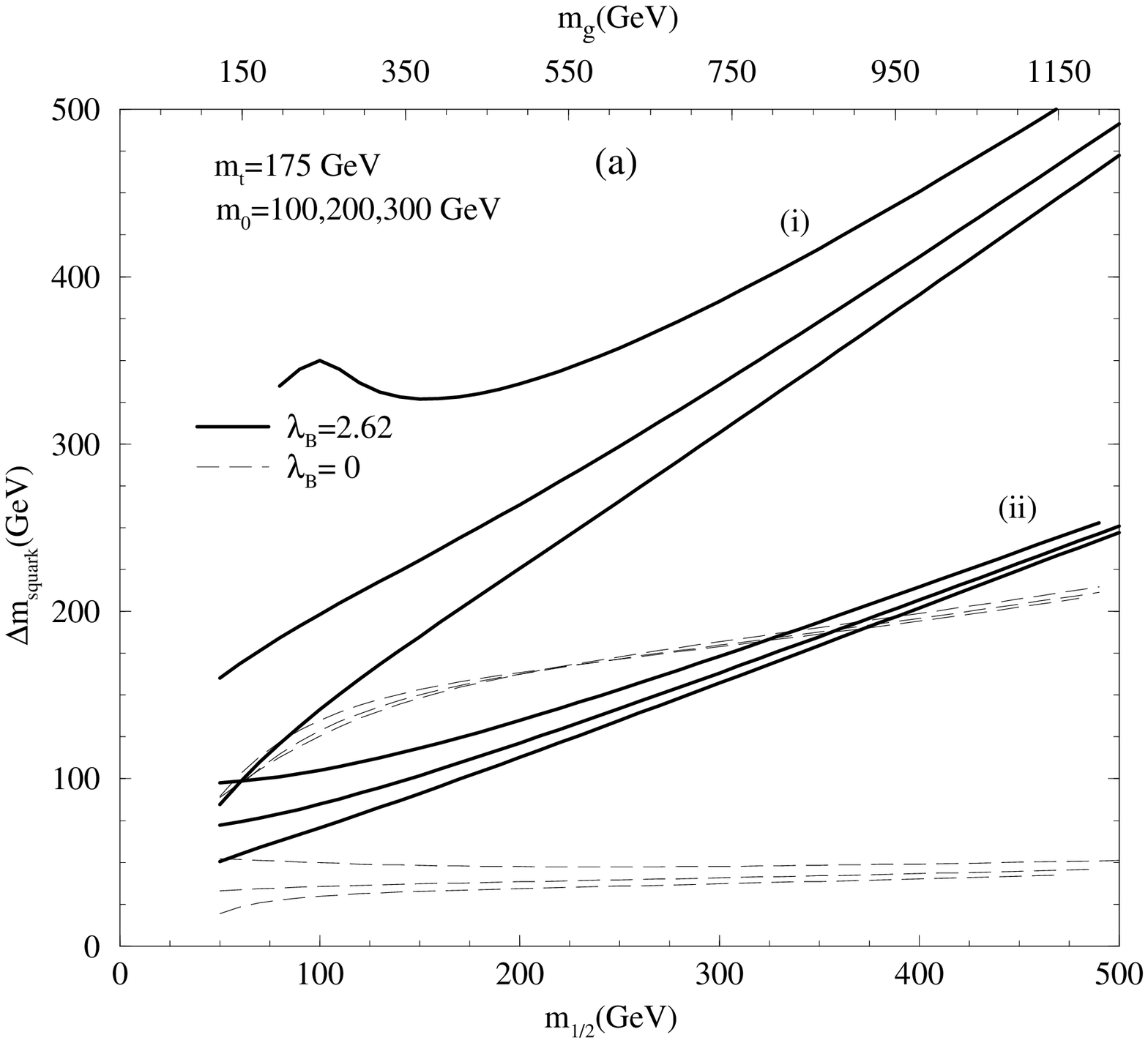} \hfil
& \epsfysize=7cm \epsfxsize=7cm \hfil \epsfbox{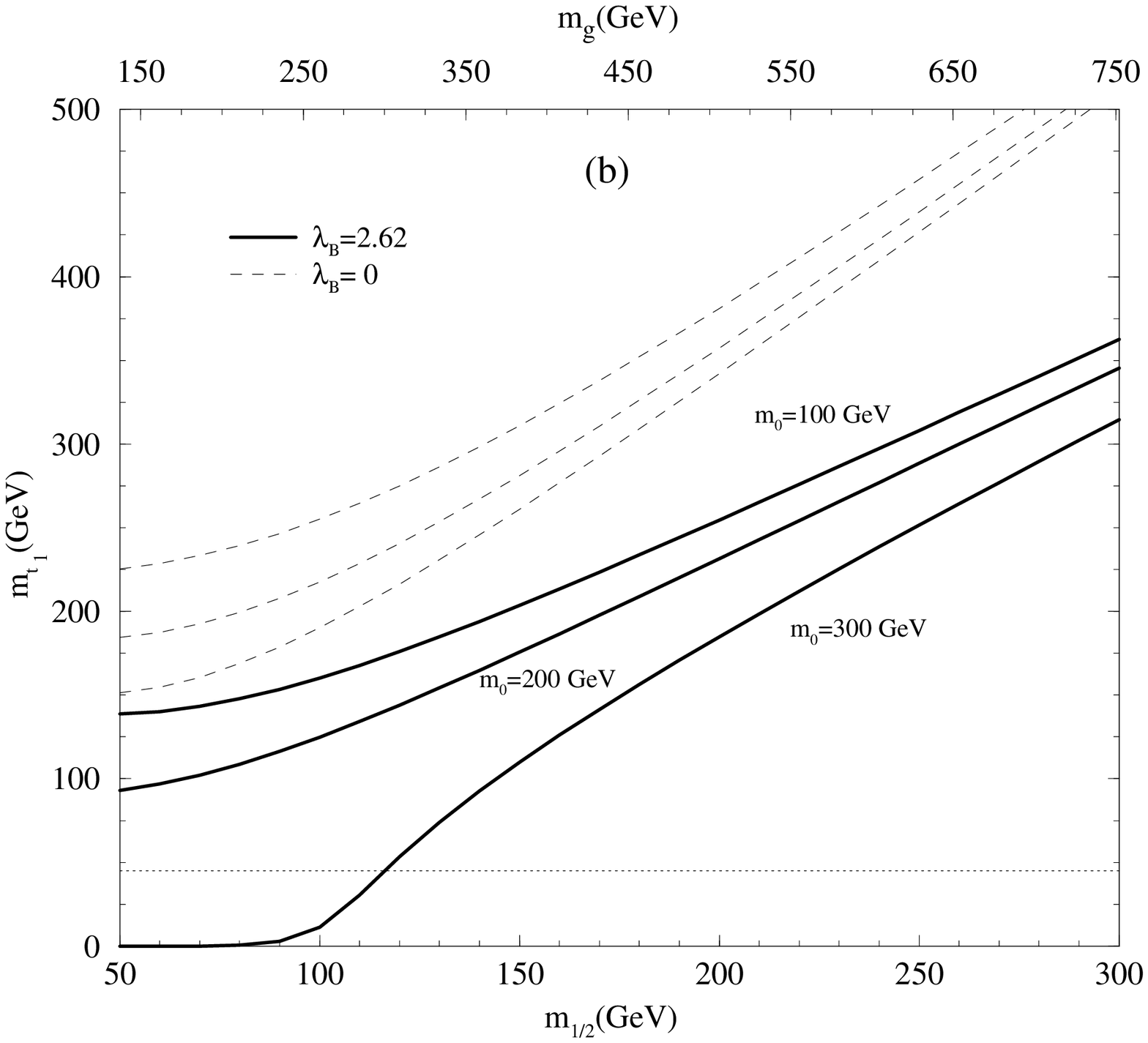} \hfil
\end{tabular}
{\small \captions{(a) The splitting in the (i) stop, (ii) sbottom, for
several values of $m_0$. In each group, the lowest curve is for the
lowest value of $m_0$ and so on.  Bold solid lines represent
$\lbv(M_X) = 2.62$, and dashed lines represent $\lbv=0$, and in both
cases $\tan\beta=14.99$.  (b) The lower eigenvalue of the stop system.
Same convention of solid and dashed lines as before. The horizontal
line is at 45 GeV, the lower bound from LEP. For $m_0=300$ GeV the
lower stop may become lighter than the LEP bound for $m_{1/2}< 120$
GeV. We note that in $\lbv =2.62$ case the lowest stop mass reduces
with increasing $m_0$, contrary to the $\lbv=0$ case.} }
\end{figure}
} { \oddsidemargin -1cm
\begin{figure}[t]
\begin{tabular}{cc}
\epsfysize=7cm \epsfxsize=7cm \hfil \epsfbox{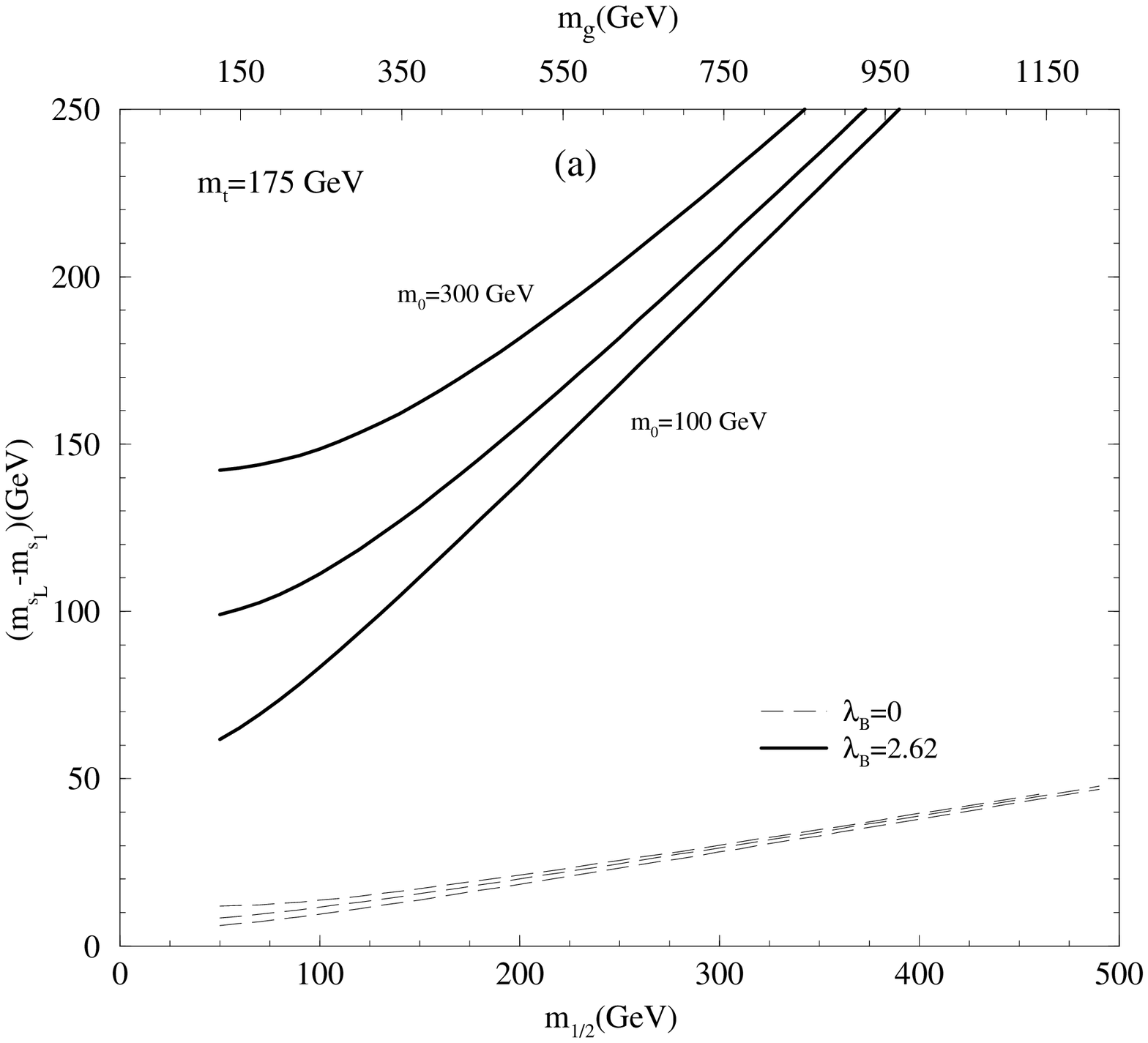} \hfil
& \epsfysize=7cm \epsfxsize=7cm \hfil \epsfbox{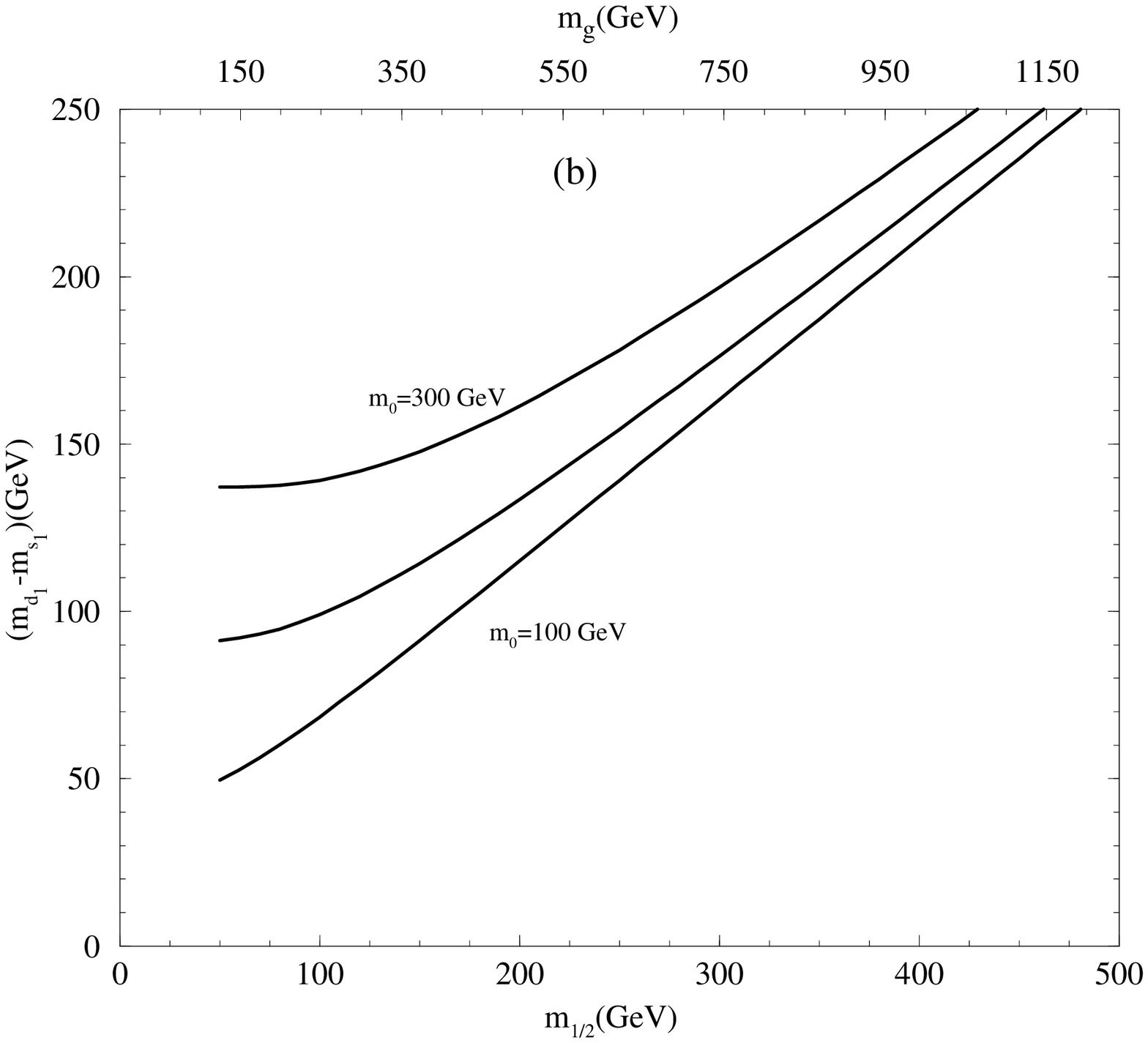} \hfil
\end{tabular}
{\small \captions{(a) Splitting between the left-handed
s-strange squark and the lighter eigenvalue of the mass matrix in Eqn.
(9), for several values of $m_0$; dashed lines are as before. 
(b) The difference in eigenvalues of the first and second generation
right handed squark mass matrix. For the analytical expression see
Eqn. (14) and Eqn. (15). Subscript 1 refers to the lighter eigenvalue
as in the stop system.} } 
\end{figure}
}

{ \doublespace
The additional splitting created between the squark eigenvalues in the
$\lbv \ne 0 $ case is mainly due to the induced splitting between the
running masses in the diagonal entries, and it can be understood if we
write down the RGEs for the differences in the left and right-handed
running mass parameters.  We have defined $Y_k=h_k^2/4 \pi$, where
$h_k$ is a relevant Yukawa coupling and $t=\ln \mu /2 \pi$, where
$\mu$ is the renormalization scale and we have used the notation
$Y_i=Y_{i33}$.  The differences in the diagonal entries renormalize
as, }

{\oddsidemargin -1cm
{ \small
\bea
{\partial(\tilde{m}^2_{t_L}-\tilde{m}^2_{t_R})\over\partial t} &=&
-\left( 3\alpha_2 M_2^2+ Y_t \Delta_u +2 Y_1 \Delta_1 + 2 Y_2 \Delta_2
+ 4 \sqrt{Y_1 Y_2} {\tilde{m}}^2_{ds}
\right)+\alpha_1 M_1^2 +Y_b \Delta_d \,, \label{splitt}
\\
{\partial(\tilde{m}^2_{b_L}-\tilde{m}^2_{b_R})\over \partial t} &=&
-\left( 3 \alpha_2 M_2^2+ Y_b \Delta_d + 2 Y_1 \Delta_1 + 2 Y_2
\Delta_2 + 4 \sqrt{Y_1 Y_2} {\tilde{m}}^2_{ds}
\right)+ \frac{1}{5}\alpha_1 M_1^2 + Y_t \Delta_u \label{splitb}\,,\\
{\partial(\tilde{m}^2_{s_L}-\tilde{m}^2_{s_R})\over \partial t} &=&
-\left( 3 \alpha_2 M_2^2+ 2 Y_2 \Delta_2 + 2 \sqrt{Y_1 Y_2}
{\tilde{m}}^2_{ds}
\right)+ \frac{1}{5}\alpha_1 M_1^2 \label{splits}
\eea
}} 
In the right handed mass matrix, the difference in eigenvalues can
be computed from,
\bea
{\partial(\tilde{m}^2_{d_R}-\tilde{m}^2_{s_R})\over \partial t} &=& 2
(Y_1 \Delta_1 - Y_2 \Delta_2)\,,\label{splitds}\\
{\partial({\tilde{m}}^2_{ds})\over \partial t}~~~~ &=& (Y_1 +Y_2
)~{\tilde{m}}^2_{ds} + \sqrt{Y_1 Y_2} ~(\Delta_1 +
\Delta_2). \label{splitofd} \eea
with,
\bea
\Delta_u &=& A^2_t + \tilde{m}^2_{q_{L_3}} + m^2_{H_2} + 
\tilde{m}^2_{u_{R_3}}\,, \nonumber \\
\Delta_d &=& A^2_b + \tilde{m}^2_{q_{L_3}} + m^2_{H_1} + 
\tilde{m}^2_{d_{R_3}}\,, \nonumber \\
\Delta_i &=& 
 A^2_{i33} + \tilde{m}^2_{u_{R_3}}+ \tilde{m}^2_{d_{R_3}} +
\tilde{m}^2_{d_{R_i}}, 
\eea
When $\lbv$ is turned on in the R.H.S. of Eqn.(\ref{splitt}),
(\ref{splitb}), (\ref{splits}) the splitting between the squarks is
increased. This also happens with increasing $m_{1/2}$ due to the
presence of $M_2$ in the right hand side. Large values of $m_{1/2}$
also increase the $\Delta$'s and this gives a indirect additional
effect to amplify the splittings. The $\Delta$'s also grow with $m_0$
and hence the splitting is also enhanced by the rise in $m_0$. The
flattening of the curves in the $\lbv=0$ case in Fig. (2.a) is due to the
off-diagonal elements in the mass matrices. In the $\lbv \ne 0$ case,
both the values of $\mu$ and $A_u$ reduce at low energy compared to
the B-conserving case making the off-diagonal effect milder. In
Eqn.(\ref{splitt}) $Y_t$ has an increasing effect (negative sign)
while $Y_b$ a decreasing effect (positive sign). Whereas in
Eqn.(\ref{splitb}) $Y_b$ and $Y_t$ exchange their roles. Because $Y_t$
is much greater than $Y_b$, splitting in the stop system is more
pronounced.
 
The absence of Flavor Changing Neutral Current (FCNC) constrains the
off-diagonal mixing terms between the first and the second generation
\cite{fcnc} denoted by ${\tilde{m}}^2_{ds}$ in Eqn. (\ref{dsmat}). 
Even though at the GUT scale the mixing is taken to be vanishing, it is
generated by one loop diagrams connecting the right handed $d$ and $s$
squarks when $\lbf$ and $\lbs$ are present simultaneously. However
this diagram is suppressed when one coupling is sufficiently smaller
than the other \cite{mar5}.  Now we will show, that the maximum
mass-splittings plotted in Figures (2) and (3) are independent of the
relative magnitude of $\lbf$ and $\lbs$.  In other words, even after
taking into account the constraints from FCNC on the ratio
$\lbf/\lbs$, the mass splittings remain unaltered for given value of
the sum $\sqrt{{\lbf}^2+{\lbs}^2}$.

Let us re-write the splitting in the stop system in terms of the
quantities,
\be
r^2={Y_1/Y_2} ~~~{\rm and} ~~~c(r)= \Delta_1 Y_1 + \Delta_2 Y_2 + 2
\sqrt{Y_1 Y_2} ~{\tilde{m}}^2_{ds}.
\ee
The Eqn. (\ref{splitt}) is of the general form,
\be
{\partial(\tilde{m}^2_{t_L}-\tilde{m}^2_{t_R})\over\partial t} = f
-2~c(r),
\ee
where, $f$ is independent of $r$ and the combination $c$ renormalize as,
\be
{\partial c(r) \over \partial t} = g + h ~c(r), \label{c}
\ee
$g$ and $h$ being independent of $r$. 
Differentiating Eqn. (\ref{c}) with respect to $r$ and performing the
integration of Eqn. (\ref{c}) from the scale $M_X$ to a scale $\mu$ we
get,
\be
\left. \left. \frac{\partial c}{\partial r} \right |_\mu= \frac{\partial 
c }{ \partial r}\right |_{M_X} e^{\int_{M_X}^{\mu} h~ dt}.
\ee
Universality of the soft masses at the scale $M_X$ gives $\left.
\frac{\partial c }{ \partial t }\right |_{M_X}=0$. 
Hence ${\partial c / \partial r}$
vanishes at all scales proving $c$ to be independent of $r$. As Eqn.
(\ref{splitt}) is independent of $r$, so is the  splitting for
a given value of $\lbv$. For the first two generations the eigenvalues
of the right-handed-squark mass matrix can be shown to be independent
of $r$ by repeated application of this procedure to the evolution of
the trace and the determinant of the mass-matrix given in
Eqn. (\ref{dsmat}).

To conclude, non zero $\lbv$ introduces asymmetric renormalization
effects between the left and right handed sector of squarks,
increasing the splitting between them, as shown in Figure (2) and (3),
which has been obtained by full numerical integration of the all soft
masses and the third generation dimensionless Yukawa couplings of MSSM
extended by including baryon non-conserving Yukawa couplings of the
third generation, and properly taking into account the constraints 
imposed by correct radiative symmetry breaking. We have shown that these 
splittings are independent
of $\lbf/\lbs$ but depends on $\lbv$.  The pattern that emerged
here is that the splitting grows with increasing $m_0$ and
$m_{1/2}$. Our solid lines for the stop system in Figure (2.a) can be
interpreted as the maximum possible splitting (corresponding to
$\lbv(M_X)=2.62$ or $\lbv(m_t)=1.03$), and similarly for the sbottom
and s-strange systems. 

The lower eigenvalue in the stop
and the sbottom system is considerably lower than the $\lbv=0$
case. The lightest stop mass is predicted to be below the experimental 
bound of 45 GeV for ranges of parameter space when $\lbv$ is large. 
Hence, when B-violating couplings are included, wider ranges of parameter 
space of the supersymmetric theory can be eliminated from the 
experimental lower bound on squark masses. Contrary to the naive 
expectations, the mass of the lightest stop $reduces$ with increasing $m_0$.

The searches of R-parity violating supersymmetry in colliders
\cite{baer} have to take into account the susy mass spectrum with 
the proper renormalization effects due to R-parity violating couplings. 
On the other hand, the splitting in the stop system would also influence 
chargino contribution to the flavor changing decay 
$b \rightarrow s+\gamma$ \cite{bsg}.
\vskip 1cm
\noindent We acknowledge comments from  F. Vissani, M. Drees and S.
Bertolini. 
\vskip 1cm 

\noindent After the completion of this work, we came across the preprint by
B. de Carlos and P. L. White, hep-ph/9602381, OUTP-96-01P, where RGE
for the soft terms including R-parity violating Yukawa couplings have
been derived. Our RGEs agree with them, however they have concentrated
on leptonic flavor violating effects induced by L violating couplings.

\end{document}